\documentclass[12pt]{article}
\usepackage{epsf}

\renewcommand{\thefootnote}{\#\arabic{footnote}}
\begin{document}

\newcommand{\gtrsim}{ \mathop{}_{\textstyle \sim}^{\textstyle >} }
\newcommand{\lesssim}{ \mathop{}_{\textstyle \sim}^{\textstyle <} }

\renewcommand{\thefootnote}{\fnsymbol{footnote}}
\setcounter{footnote}{0}
\begin{titlepage}

\def\thefootnote{\fnsymbol{footnote}}

\begin{center}

\hfill RESCEU-04/01\\
\hfill TU-622\\
\hfill astro-ph/0105161\\
\hfill May, 2001\\
\vskip .5in

{\Large \bf
Cosmic Microwave Background Anisotropy\\ 
with Cosine-Type Quintessence
}

\vskip .45in

{\large
Masahiro Kawasaki$^{(a)}$,
Takeo Moroi$^{(b)}$ and Tomo Takahashi$^{(b)}$
}

\vskip .45in

{\em
$^{(a)}$Research Center for the Early Universe, School of Science,
University of Tokyo\\
Tokyo 113-0033, Japan
}
\vskip .2in

{\em
$^{(b)}$Department of Physics,  Tohoku University\\
Sendai 980-8578, Japan
}

\end{center}

\vskip .4in

\begin{abstract}

    We study the Cosmic Microwave Background (CMB) anisotropies
    produced by cosine-type quintessence models.  In our analysis,
    effects of the adiabatic and isocurvature fluctuations are both
    taken into account. For purely adiabatic fluctuations with scale
    invariant spectrum, we obtain a stringent constraint on the model
    parameters using the CMB data from COBE, BOOMERanG and MAXIMA.
    Furthermore, it is shown that isocurvature fluctuations have
    significant effects on the CMB angular power spectrum at low
    multipoles in some parameter space, which may be detectable in
    future satellite experiments.  Such a signal may be used to test
    the cosine-type quintessence models.

\end{abstract}
\end{titlepage}

\renewcommand{\thepage}{\arabic{page}}
\setcounter{page}{1}
\renewcommand{\thefootnote}{\#\arabic{footnote}}
\setcounter{footnote}{0}

\section{Introduction}
\label{sec:intro}
\setcounter{equation}{0}

It is widely believed that the present cosmological observations are
consistent with low density Cold Dark Matter (CDM) models with scale
invariant adiabatic density fluctuations.  The required energy density
of non-relativistic matter is about $30-40$\% of the critical density
(i.e., $\Omega_{\rm m}\simeq 0.3-0.4$).  On the other hand, the recent
BOOMERanG \cite{boomerang} and MAXIMA \cite{maxima} experiments on
anisotropies of Cosmic Microwave Background (CMB) strongly suggest
that our universe is flat, which is also the prediction of the
inflationary universe.  Thus, there exists dark energy which fills the
gap between $\Omega_{\rm m}$ and the total energy density $\Omega_{\rm
tot}$. Although usually the dark energy is assumed to be the
cosmological constant, a slowly evolving scalar field with positive
energy can also account for the dark energy \cite{tracker,cosine}.
Such a scalar field is called quintessence and has been studied by
many authors \cite{quintessence}.

At present, the quintessence models are classified into two types;
tracker type \cite{tracker} and cosine type \cite{cosine}.  The former
has an attractor-like solution which explains the present dark energy
without fine-tuning of the initial condition, while the latter needs
tuning of the initial value of the scalar field.  This is why the
tracker type is favored among cosmologists.  However, even for the
tracker-type models the model parameters should be fine-tuned to
produce the required value of the present quintessence density.  In
addition, the potential for the tracker field is very exotic and hard
to realize in particle physics models.\footnote{
For model building of the tracker field, see Refs.~\cite{Masiero}.}

One of the observational effects produced by the existence of the
quintessence is the CMB anisotropies.  In many cases, the quintessence
dominates the universe at late times ($z \sim O(1)$, with $z$ being
redshift) after the recombination.  At that epoch the gravitational
potential is changed because the equation of state for quintessence is
different from that for non-relativistic matter, which leads to an
enhancement of the CMB anisotropies at large angular scales $l
\lesssim 10$ (where $l$ is the index of spherical harmonics) due to
the late-time integrated Sachs-Wolfe effect \cite{SacWol}.  The
quintessence also changes the locations of the acoustic peaks in the
CMB angular power spectrum \cite{Brax} because the projection of the
horizon at last scattering onto the present sky is enlarged compared
with models with the cosmological constant.

Furthermore, in the case of cosine-type models, during inflation the
quintessence scalar field is effectively massless and has significant
quantum fluctuations as large as $H_{\rm inf}/2\pi$ where $H_{\rm
inf}$ is the Hubble parameter during inflation.  These fluctuations
behave as isocurvature mode and hence the quintessence may have both
adiabatic and isocurvature perturbations. The isocurvature
fluctuations produce adiabatic ones once their wavelengths enter the
horizon.  Therefore, we expect larger anisotropies in CMB angular
spectrum for cosine-type quintessence models.\footnote{
Abramo and Finelli~\cite{Abramo} studied the isocurvature fluctuations
for tracker-type quintessence models without considering the evolution
of the tracker field and its fluctuations during inflation.}

In this paper we study the CMB anisotropies produced by quintessence
models paying attention to the cosine-type quintessence models. It is
shown that the isocurvature fluctuations have significant effects on
the CMB angular power spectrum at low multipoles in some parameter
space, which may be detectable in future satellite experiments and can
be an interesting signal of the cosine-type quintessence models.

This paper is organized as follows.  In Section \ref{sec:zeromode}, we
discuss dynamics of the scalar-field zero mode. In Section
\ref{sec:fluctuation}, behaviors of the fluctuation in the
quintessence amplitude are discussed.  Then, in Section
\ref{sec:numerical}, CMB anisotropy in cosmological models with
quintessence is numerically calculated.  Section \ref{sec:conclusion}
is devoted for conclusions and discussion.

\section{Dynamics of Scalar-Field Zero Mode}
\label{sec:zeromode}
\setcounter{equation}{0}

We start our discussion with the behavior of the zero mode of the
quintessence field $Q(t,\vec{x})$.  We denote the zero mode of the
quintessence field as $\bar{Q}(t)$.

The energy-momentum tensor of the quintessence field is given by
\begin{eqnarray}
    T_{\mu\nu} = \partial_\mu Q \partial_\nu Q
    - \frac{1}{2}
    \left[ \partial^\alpha Q \partial_\alpha Q + 2 V(Q) \right]
    g_{\mu\nu},
\end{eqnarray}
where $V$ is the quintessence potential.  Thus, for $Q=\bar{Q}$, the
energy density and the pressure of the quintessence field are
\begin{eqnarray}
    \rho_Q = \frac{1}{2} \dot{\bar{Q}}^2 + V(\bar{Q}),~~~
     p_Q = \frac{1}{2} \dot{\bar{Q}}^2 - V(\bar{Q}),
\end{eqnarray}
respectively, where the ``dot'' represents the derivative with respect
to time $t$.  If the kinetic energy of $Q$ is negligible compared to
the potential energy, the equation of state becomes $\omega_Q\equiv
p_Q /\rho_Q\simeq -1$ and the energy-momentum tensor of the
quintessence field behaves like that of the cosmological constant.

The zero mode $\bar{Q}$ obeys the following equation of motion
\begin{eqnarray}
    \ddot{\bar{Q}} + 3 H \dot{\bar{Q}} + V'(\bar{Q}) = 0,
\end{eqnarray}
where the ``prime'' denotes the derivative with respect to $Q$.  In
the very early universe where $H^2\gg |V'/\bar{Q}|$, the slow-roll
condition is satisfied and the motion of $\bar{Q}$ is negligible even
if $\bar{Q}$ is displaced from the minimum of the potential.  If the
slow-roll condition is satisfied until today, the equation of state
$\omega_Q$ is always very close to $-1$ and the quintessence is
indistinguishable from the cosmological constant.  However, this is
not always the case.  If the effective mass of the quintessence may
become larger than the expansion rate, $Q$ starts to move and
$\omega_Q$ varies.  In addition, since the quintessence is a dynamical
field, its energy density may fluctuate.  Due to these facts,
quintessence models may have an interesting consequence in the
evolution of the cosmological perturbations, in particular, in the CMB
anisotropy.  The epoch when the quintessence starts to move is
strongly model-dependent.  If the slow-roll condition is satisfied
until very recently, the quintessence field may provide significant
amount of the energy density with the equation of state close to $-1$
to the total energy density of the universe.  This may be a solution
to the ``dark energy'' problem alternative to the cosmological
constant.

In this paper, we consider the cosine-type quintessence potential:
\begin{eqnarray}
     V(Q) = \Lambda^4
     \left[ 1 -
         \cos\left(\frac{Q}{f_Q}\right) \right]
     = 2\Lambda^4
     \sin^2\left(\frac{Q}{2f_Q}\right).
     \label{V_q}
\end{eqnarray}
This type of potential can be generated if the quintessence field is a
pseudo Nambu-Goldstone boson.  In this class of models, effective mass
of the quintessence field is always of $O(\Lambda^2/f_Q)$, and is
insensitive to the amplitude $\bar{Q}$.  Requiring that the slow-roll
condition for the quintessence field be satisfied until very recently
to realize $\omega_Q\simeq -1$, the combination $\Lambda^2/f_Q$ cannot
be much larger than the present expansion rate of the universe.
Consequently, motion of the quintessence field is negligible for $z\gg
1$.  Thus, in this class of models, the present energy density of the
quintessence field sensitively depends on the initial amplitude.

In the cosine-type models, the present density parameter for the
quintessence depends on the following three parameters: $f_Q$,
$\Lambda$, and the initial amplitude of the quintessence field denoted
as $\bar{Q}_{\rm I}$.  Assuming the flat universe (i.e., $\Omega_{\rm
m}+\Omega_Q=1$), however, one relation among these parameters holds
once we fix the present matter density $\Omega_{\rm m}=\Omega_{\rm
CDM}+\Omega_{\rm b}$, where $\Omega_{\rm CDM}$ and $\Omega_{\rm b}$
are density parameters for the CDM and baryons, respectively.

\begin{figure}[t]
    \centerline{\epsfxsize=0.75\textwidth\epsfbox{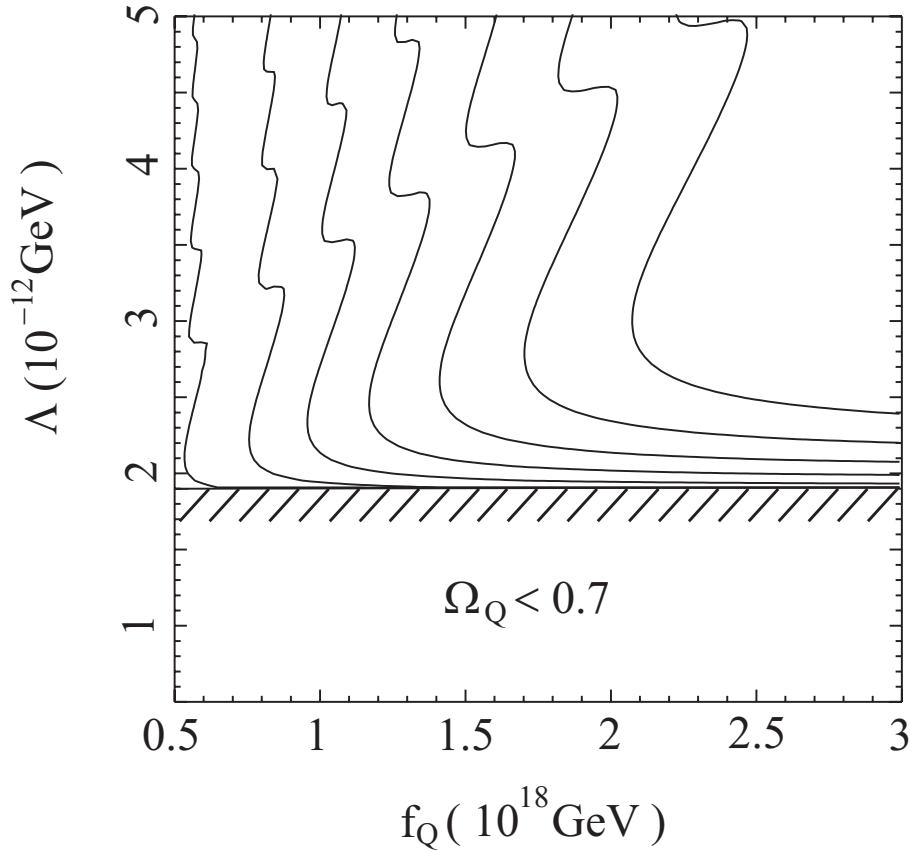}}
    \caption{Contours of constant $\bar{Q}_{\rm I}/ f_{Q} 
      = 3.0, 2.75, 2.5, 2.25, 2.0, 1.75$ and 1.5 from left to right.
      The value of $\bar{Q}_{\rm I}$ is determined  such that 
      $\Omega_{\rm  Q}=0.7$ in the flat universe (i.e., 
      $\Omega_{\rm  m}+\Omega_Q=1$) today.  
      The Hubble parameter is taken to be
      $h=0.65$. The parameter space $\Lambda \lesssim 1.9 \times
      10^{-12}$ GeV is the region where $\Omega_{\rm Q}=0.7$ cannot be
      realized for any values of $\bar{Q}_{\rm I}$.}
    \label{fig:phi_I}
\end{figure}

In Fig.\ \ref{fig:phi_I}, we plot the contours of constant
$\bar{Q}_{\rm I}$ which realizes $\Omega_Q=0.7$ in the flat universe.
Here, we take $h=0.65$, where $h$ is the present Hubble parameter in
units of 100 km/sec/Mpc.  Notice that there is no possible value of
$\bar{Q}_{\rm I}$ consistent with $\Omega_Q=0.7$ for $\Lambda\lesssim
1.9 \times 10^{-12} {\rm GeV}$.  This is because the energy density of
the quintessence is at most $2\Lambda^4$ and hence $\rho_Q$ cannot be
large enough if $2\Lambda^4<\Omega_Q\rho_{\rm c}$ with $\rho_{\rm c}$
being the critical density of the present universe. When $2\Lambda^4$
is close to $\Omega_Q\rho_{\rm c}$, $\bar{Q}_{\rm I}$ should be close
to $\pi f_Q$ to realize the relevant value of $\Omega_Q$.  In this
case, motion of the quintessence field is almost negligible and energy
density of the quintessence behaves like that of the cosmological
constant.

\begin{figure}[t]
   \centerline{\epsfxsize=0.75\textwidth\epsfbox{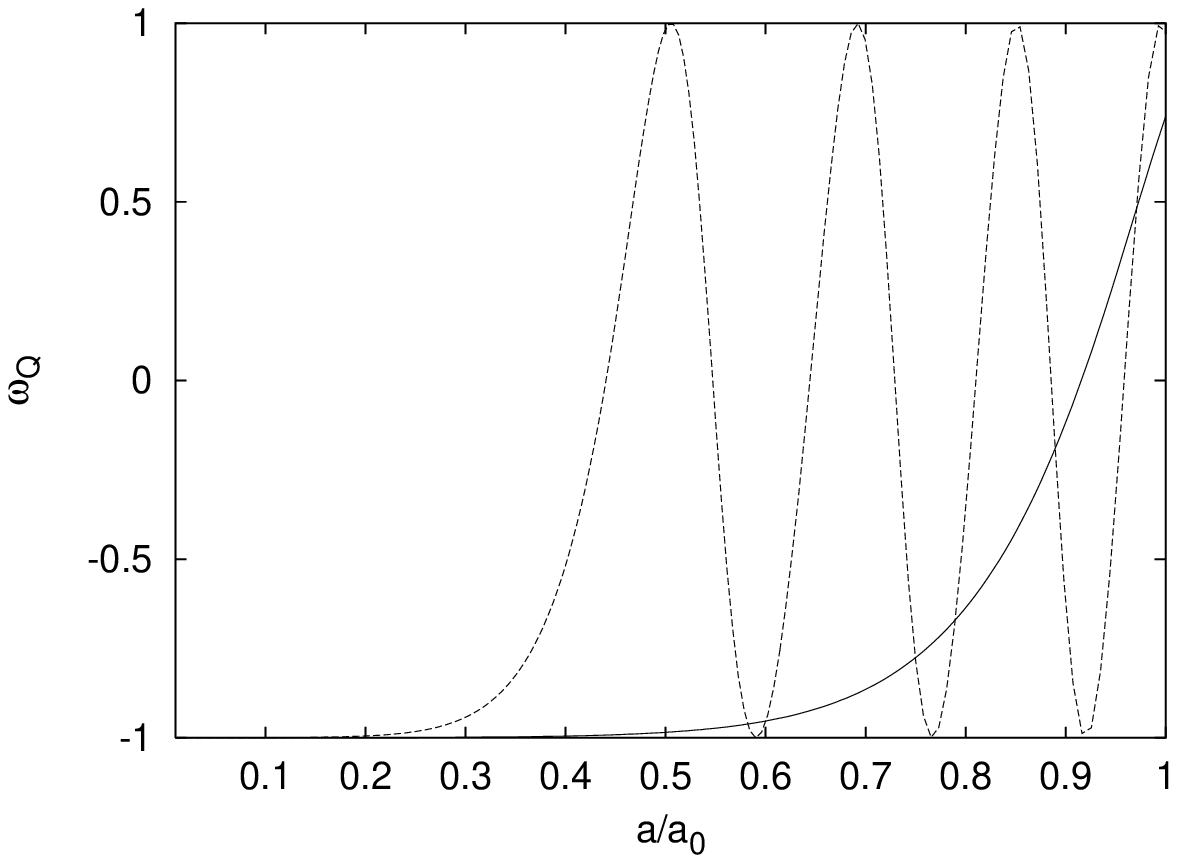}}
   \caption{Evolution of the equation of state $\omega_Q$ for (a)
   $f_Q=5.0 \times 10^{17}$ GeV and $\Lambda=2.3 \times 10^{-12}$ GeV
   (solid line), and (b) $f_Q=5.0 \times 10^{17}$ GeV and $\Lambda=4.0
   \times 10^{-12}$ GeV (dashed line).  Here, we take $h=0.65$,
   $\Omega_{\rm m}=0.3$ and $\Omega_Q=0.7$.}
   \label{fig:EqState}
\end{figure}

On the contrary, for larger $\Lambda$, $\bar{Q}_{\rm I}$ smaller than
$\pi f_Q$ is possible and the motion of the quintessence becomes
important.  In particular, if $\Lambda$ becomes large enough, the
quintessence oscillates around the minimum of the potential.  As one
can see, for large enough $\Lambda$, the contours have an oscillatory
behavior.  This can be understood as follows.  When the relation
$H\lesssim\Lambda^2/f_Q$ is satisfied, the quintessence field starts
to oscillate.  If the combination $\Lambda^2/f_Q$ is large, the
oscillation starts earlier epoch and the quintessence field undergoes
many oscillations until the present time.  We can read off this
behavior from Fig.\ \ref{fig:phi_I}.  We call the parameter space
where the oscillation of the quintessence is significant as
``oscillatory region.''  When the quintessence field starts to
oscillate earlier, it dominates the energy density of the universe
from earlier epoch.  This has significant implications to the CMB
power spectrum.

Evolution of $\omega_Q$ is also important since it can be quite
different from the case of the cosmological constant, in particular
when the motion of the quintessence is non-negligible.  Typical
behaviors of $\omega_Q$ as a function of the scale factor are shown in
Fig.\ \ref{fig:EqState}.  When the curvature of the potential is
smaller than the expansion rate of the universe, evolution of the
quintessence can be neglected, and $\omega_Q\simeq -1$.  On the
contrary, when $H\lesssim\Lambda^2/f_Q$, $\omega_Q$ oscillates between
$-1$ and $+1$.  If the oscillation is fast enough, $\omega_Q$ is
effectively $0$ taking the average over the oscillation.  In this
case, the scalar field $Q$ behaves as a non-relativistic matter.  If
the condition $H\sim\Lambda^2/f_Q$ is realized in a very recent
universe, however, the situation is quite different.  In this case,
the quintessence field starts to move very recently and the averaged
equation of state can be smaller than $0$.  If so, the scalar field
$Q$ behaves differently from non-relativistic matter and cosmological
constant.  This may have interesting impacts on the CMB anisotropy as
we will see in the following sections.

\begin{figure}[t]
   \centerline{\epsfxsize=0.75\textwidth\epsfbox{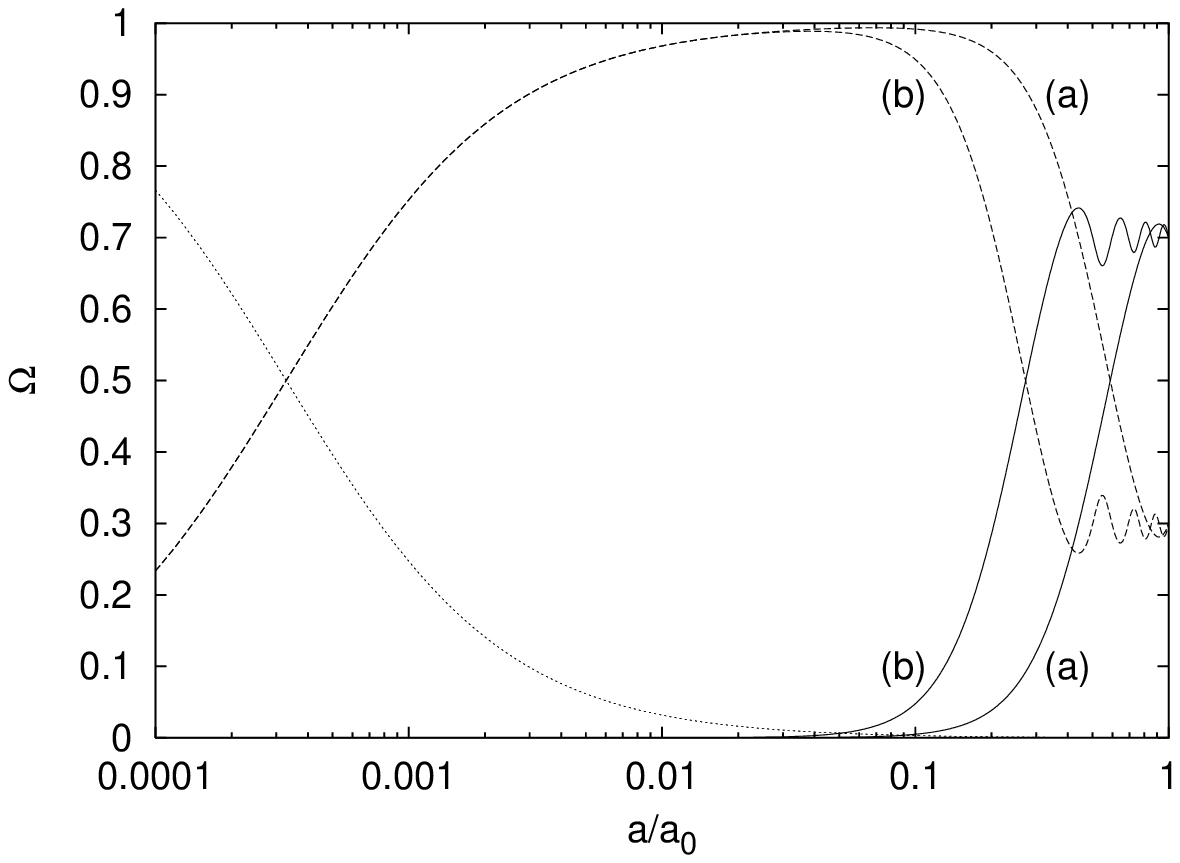}}
   \caption{Evolution of the density parameters. We show 
   $\Omega_Q$ (solid line), $\Omega_{{\rm m}}$ (dashed line) 
   and $\Omega_{\rm radiation}$ (dotted line). Cosmological and model 
   parameters are the same as Fig.\ \ref{fig:EqState}}
   \label{fig:Omega_Q}
\end{figure}

We also present the evolution of the density parameter of the
quintessence:
\begin{eqnarray}
   \Omega_Q (t) \equiv \frac{\rho_Q(t)}{\rho_{\rm tot} (t)}.
   \label{OmegaQ(t)}
\end{eqnarray}
Like the case of the cosmological constant, $\Omega_Q (t)$ is
negligibly small in the early universe.  When $z\sim O(1)$, however,
$\Omega_Q (t)$ becomes close to 1 and the energy density of the
quintessence becomes sizable.  One interesting feature is that, if the
slow-roll condition does not hold, the behavior of $\Omega_Q (t)$ is
quite different from that of the cosmological constant.  In
particular, as $\Lambda$ becomes larger, $\rho_Q$ becomes the dominant
component of the total energy density at earlier stage.

The observations of Type Ia Supernovae suggest that $\omega_{Q}$ is
less than about $-0.7$ at $z\lesssim 1$ for $\Omega_{\rm m}=0.3$
\cite{Perlmutter}.  However, it was pointed that there exist some
possible systematic errors such as dust absorption and/or evolution
effect \cite{Dominguez}.  Therefore, in this paper, we do not use the
supernovae data to obtain constraints on the quintessence models.

\section{Fluctuation in the Quintessence Field}
\label{sec:fluctuation}
\setcounter{equation}{0}

Since the quintessence is a scalar field, its amplitude may have
position-dependent fluctuations.  To investigate its behavior, we
decompose the $Q$ field as
\begin{eqnarray}
    Q (t, \vec{x}) = \bar{Q} (t) + q (t, \vec{x}),
\end{eqnarray}
where $q$ is the perturbation of the amplitude of the quintessence
field.  Hereafter, we study the evolution of $q$ using the linearized
equations for the perturbations.

The equation of motion for $q$ is, in the synchronous gauge,
\begin{eqnarray}
    \ddot{q} + 3 H \dot{q} - \left( \frac{a}{a_0} \right)^{-2} 
    \partial_i^2 q
    + V''(\bar{Q}) q = -\frac{1}{2} \dot{h} \dot{\bar{Q}},
    \label{eom_dq}
\end{eqnarray}
where $\partial_i$ is the derivative with respect to the comoving
coordinate $x^i$.  Here, the perturbed line element in the synchronous
gauge is given by
\begin{eqnarray}
    ds^2 = -dt^2 + \left( \frac{a}{a_0} \right)^2
    (\delta_{ij} + h_{ij}) dx^idx^j
    =   \left( \frac{a}{a_0} \right)^2
    \left[ -d\tau^2 + (\delta_{ij} + h_{ij}) dx^idx^j
    \right],
\end{eqnarray}
where $\tau$ is the conformal time coordinate, $a$ the scale factor at
time $t$, $a_0$ the scale factor at the present time, and $h$ is the
trace of $h_{ij}$.\footnote{
The metric perturbation $h$ should not be confused with the Hubble
parameter.}
In the momentum space, we expand
$h_{ij}$ as
\begin{eqnarray}
    h_{ij} (t, \vec{x}) = \int \frac{d^3k}{(2\pi)^{3/2}}
    \left[ \frac{k_ik_j}{k^2} h(\vec{k},t)
    + \left( \frac{k_ik_j}{k^2} - \frac{1}{3}\delta_{ij} \right)
    6 \eta (\vec{k},t) \right] e^{i\vec{k}\vec{x}},
    \label{h_ij}
\end{eqnarray}
with $k^2=k_ik^i$.  Notice that, in the momentum space, the
gauge-invariant variable $\Psi$ is related to $h$ and $\eta$ as
\begin{eqnarray}
    \Psi (k) = \frac{1}{2k^2} \left[
    \frac{\partial^2 h (k)}{\partial\tau^2}
    + 6\frac{\partial^2 \eta (k)}{\partial\tau^2}
    + \frac{1}{a} \frac{\partial a}{\partial\tau}
    \left( \frac{\partial h (k)}{\partial\tau}
    + 6\frac{\partial \eta (k)}{\partial\tau} \right)
    \right].
\end{eqnarray}
Numerically, we checked that the terms proportional to the derivatives
of $\eta$ are subdominant relative to contributions from the
derivatives of $h$.

There are two origins of non-vanishing $q$ at the present time. As
indicated in Eq.\ (\ref{eom_dq}), the metric perturbation $h$
generates non-vanishing $q$ even if $q$ initially vanishes.  We call
this fluctuation as ``adiabatic fluctuation'' of the quintessence
field.  Let us briefly discuss how the adiabatic fluctuation of $q$ is
generated.  For this purpose, it is instructive to consider the
matter-dominated universe even though, in our case, the energy density
of the present universe has significant contribution from the
quintessence.  Fluctuation for the scale smaller than the horizon
scale oscillates and damps, so we consider $q(k)$ which has larger
physical momentum than the expansion rate of the universe.  Then, the
third term in the left-hand side of Eq.\ (\ref{eom_dq}), i.e.,
$\partial_i^2q$, is negligible.  In solving Eq.\ (\ref{eom_dq}), it is
convenient to use the fact that $\Psi$ is (almost) constant of time in
matter-dominated epoch.  In addition, for the zero mode, we use the
slow-roll condition to derive $\dot{\bar{Q}}\simeq -
\frac{2}{9}(V'/H)$.  With these informations, $q(k)$ at a given time
$t$ is, in the synchronous gauge,
\begin{eqnarray}
    q(k) = \frac{3}{88} (a_0/a)^2 k^2 V' \Psi(k) t^{4},
\end{eqnarray}
which is proportional to $t^{8/3}$.  In the realistic situation,
however, some of the approximations used to derive the above relation
may fail.  Therefore, in our analysis, we numerically solve the
equation of motion for the quintessence field to obtain $q(k)$.

The second origin is the primordial perturbation in the quintessence
amplitude generated in the very early universe, probably during the
inflation.  We call this fluctuation as ``isocurvature fluctuation,''
since the total density fluctuation and the potential $\Psi$ vanish as
$a\rightarrow 0$ if this is the only source of the fluctuation in the
early universe.

The isocurvature mode may arise due to the quantum fluctuation during
the inflation.  In order to calculate the expected fluctuation
generated during the inflation, we quantize the scalar field in the de
Sitter background.  Identifying $q$ as a field operator, it can be
expanded as
\begin{eqnarray}
    q (t,\vec{x}) = \int \frac{d^3k}{(2\pi)^{3/2}}
    \left[ \hat{a}_k \varphi_k  (\tau) e^{i\vec{k}\vec{x}}
    + \hat{a}_k^\dagger \varphi_k^\dagger (\tau)
    e^{-i\vec{k}\vec{x}} \right],
\end{eqnarray}
where, assuming minimally coupled Lagrangian for the quintessence
field, the mode function is given by
\begin{eqnarray}
    \varphi_k (\tau) =
    \frac{\sqrt{\pi}}{2} H_{\rm inf} \tau^{3/2}
    H_\nu^{(1)} (k\tau),
\end{eqnarray}
with $H_\nu^{(1)}$ being the Hankel function of the first kind and
\begin{eqnarray}
    \nu^2 = \frac{9}{4} - \frac{m_q^2}{H_{\rm inf}^2}.
\end{eqnarray}
Here, $m_q$ is an effective mass for the quintessence field during the
inflation, and $H_{\rm inf}$ is the expansion rate of the de Sitter
background.

Adopting the Bunch-Davis vacuum~\cite{Bunch}, the operators
$\hat{a}_k$ and $\hat{a}_k^\dagger$ are identified as the creation and
annihilation operators, respectively, and satisfy the following
commutation relations:
\begin{eqnarray}
    [ \hat{a}_k, \hat{a}_l ] =
    [ \hat{a}_k^\dagger, \hat{a}_l^\dagger ] = 0,~~~
    [ \hat{a}_k, \hat{a}_l^\dagger ] =
    \delta^{(3)} (\vec{k} - \vec{l}).
\end{eqnarray}

Using these relations, the equal-time two-point function for $q$ is
given as
\begin{eqnarray}
     G(t,\vec{x} ; t, \vec{y}) &\equiv&
     \langle 0 | q (t,\vec{x}) q (t, \vec{y}) | 0 \rangle
     \nonumber \\ &=&
     \frac{\pi}{4H_{\rm inf}} \int \frac{d^3 k_{\rm phys}}{(2\pi)^3}
     \left| H_\nu^{(1)} (k_{\rm phys} / H_{\rm inf}) \right|^2
     e^{i \vec{k}_{\rm phys} (\vec{x}_{\rm phys}-\vec{y}_{\rm phys})},
\end{eqnarray}
where physical momentum $k_{\rm phys}$ is related to the comoving one
$k$ as
\begin{eqnarray}
    k_{\rm phys} = (a_0/a) k,
\end{eqnarray}
and $\vec{x}_{\rm phys}=(a/a_0)\vec{x}$.  During the inflation, the
present horizon scale is far outside of the horizon, and hence we are
interested in the behavior of $G(t,\vec{x};t,\vec{y})$ with
$|\vec{x}_{\rm phys}-\vec{y}_{\rm phys}|\gg H_{\rm inf}^{-1}$. In
other words, we only need an information of the integrand with $k_{\rm
phys}\ll H_{\rm inf}$.  Denoting
\begin{eqnarray}
     G(t,\vec{x} ; t, \vec{y}) = 
     \int \frac{d k_{\rm phys}}{k_{\rm phys}}
     |\tilde{q}(k)|^2
     e^{i \vec{k}_{\rm phys} (\vec{x}_{\rm phys}-\vec{y}_{\rm phys})},
\end{eqnarray}
$\tilde{q}(k)$ for small $k$ is given by
\begin{eqnarray}
     \tilde{q} (k) =
     \frac{2\sqrt{\pi}}{\Gamma (-\nu+1) \sin\nu\pi}
     \left( \frac{k_{\rm phys}}{2H_{\rm inf}} \right)^{-(\nu-3/2)}
     \frac{H_{\rm inf}}{2\pi}.
\end{eqnarray}
For the case where $m_q\gtrsim H_{\rm inf}$, $|(k_{\rm phys}/2H_{\rm
inf})^{-(\nu-3/2)}|\ll 1$ since we are interested in modes with
$k_{\rm phys}\ll H_{\rm inf}$.  (Notice that, in most of the inflation
models, the COBE scale corresponds to $\ln (H_{\rm inf}/k_{\rm
phys})\sim 50-60$.)  Thus, for the quintessence models with relatively
large (effective) mass during the inflation, the primordial
fluctuation in $\tilde{q}$ is extremely suppressed relative to $H_{\rm
inf}$.  If the mass of the quintessence is light, however, $\nu$
becomes close to $3/2$ and the suppression factor may become
negligible.  In this limit, $\tilde{q}(k)$ is given by
\begin{eqnarray}
     \tilde{q}(k) \simeq
     \left( \frac{k_{\rm phys}}{2H_{\rm inf}}
     \right)^{2m_q^2/3H_{\rm inf}^2}
     \frac{H_{\rm inf}}{2\pi},
     \label{q(k)}
\end{eqnarray}
and hence $\tilde{q}\simeq H_{\rm inf}/2\pi$ for $m_q\ll H_{\rm inf}$.
In order to avoid a significant suppression for the primordial
perturbation, $m_q/H_{\rm inf}\lesssim 0.1-0.2$ is enough.  In
addition, notice that, for $m_q/H_{\rm inf}\ll 1$, the resultant
$\tilde{q}$ is expected to be almost scale independent; the only
source of the scale dependence is a minor variation of the expansion
rate $H_{\rm inf}$ during the inflation.  In the following discussion,
we neglect the scale dependence of $\tilde{q}$. 

It is convenient to consider the ratio of the primordial value of
$\tilde{q}$ to that of the gauge-invariant variable $\tilde{\Psi}$ at
the radiation-dominated universe:\footnote{
In our analysis, we only consider the case with scale-independent
primordial fluctuation, and $r_q$ is treated as a scale-invariant
quantity.  When $r_q$ has a scale dependence via $\tilde{\Psi}$ and/or
$\tilde{q}$, $r_q(k)$ for the present horizon scale becomes the most
important since the quadrupole anisotropy is most strongly affected by
the isocurvature mode as will be shown below.}
\begin{eqnarray}
     r_q \equiv
     \frac{\tilde{q}}{M_*\tilde{\Psi}},
     \label{q/Psi}
\end{eqnarray}
where $M_*\simeq 2.4\times 10^{18}\ {\rm GeV}$ is the reduced Planck
scale, and $\tilde{\Psi}$ is defined as
$\langle\Psi(\vec{x})\Psi(\vec{y})\rangle=\int d\ln k
|\tilde{\Psi}(k)|^2 e^{i\vec{k}(\vec{x}-\vec{y})}$.

Solving the dynamics of the inflaton during the inflation, we obtain
$\tilde{\Psi}$ at the radiation-dominated era \cite{PRD28-629}
\begin{eqnarray}
     \tilde{\Psi} = \frac{4}{9}
     \left( \frac{H_{\rm inf}^2}{2\pi |\dot{\chi}|} \right),
\end{eqnarray}
where $\chi$ is the inflaton field.  Using the slow-roll condition for
$\chi$, and also using $\tilde{q}\simeq H_{\rm inf}/2\pi$, we obtain
\begin{eqnarray}
     r_q \simeq \frac{9}{4}\frac{M_*V'_{\rm inf}}{V_{\rm inf}},
\end{eqnarray}
where $V_{\rm inf}$ is the inflaton potential and $V'_{\rm inf}$ is
its derivative with respect to the inflaton field.\footnote{
If $f_Q$ during the inflation, which is denoted as $f_Q^{\rm (inf)}$,
is smaller than the present value, $r_q$ is enhanced by the factor
$f_Q/f_Q^{\rm (inf)}$.  This may happen if the quintessence is a
pseudo Nambu-Goldstone boson; in that case $f_Q$ is given by a vacuum
expectation value of a (real) scalar field, which varies if the scalar
potential is deformed during the inflation.}
For example, for the chaotic inflation with $V_{\rm inf}\propto\chi^p$
with $p$ being an integer, $r_q$ is given by
\begin{eqnarray}
    r_q|_{\rm chaotic} = \frac{9p M_*}{4 \chi (k_{\rm COBE})},
\end{eqnarray}
where $\chi (k_{\rm COBE})$ represents the inflaton amplitude at the
time when the COBE scale crosses the horizon.  Numerically, we found
$r_q\simeq 0.3$ $-$ 0.6 for $p=2$ $-$ 10.  Since the ratio $r_q$
generically depends on the model of the inflation, we treat $r_q$ as a
free parameter in our analysis.

Notice that, in our analysis, we treat $\tilde{\Psi}$ and the
fluctuation in the quintessence amplitude as random Gaussian
variables.  In particular, no correlation is assumed among these
quantities.  Thus, $r_q$ given in Eq.\ (\ref{q/Psi}) should be
understood as the ratio of the  expectation values of $\tilde{q}$
and $\tilde{\Psi}$, and we add adiabatic and isocurvature
contributions in quadrature in calculating the CMB anisotropy.

Now, we consider the evolution of the isocurvature fluctuation after
the inflation.  The evolution of the isocurvature mode is well
described by the equation Eq.\ (\ref{eom_dq}) with neglecting the
right-hand side (i.e., $h=0$).  This is because, when the energy
density of the quintessence is negligible, the metric perturbation $h$
is insensitive to the fluctuation in the quintessence and we can
neglect the right-hand side of Eq.\ (\ref{eom_dq}) in studying the
isocurvature perturbation.  In the early universe, this is the case
for the cosine-type quintessence.  Let us first consider the case with
large wavelength (i.e., $k_{\rm phys}\ll H$).  In this case, the third
term in Eq.\ (\ref{eom_dq}) is irrelevant.  By neglecting the
right-hand side of Eq.\ (\ref{eom_dq}), $q$ behaves like a harmonic
oscillator with effective mass-squared $V''(\bar{Q})$ in the expanding
background.  Then, we can imagine two typical cases.  If the effective
mass is much smaller than the expansion rate of the universe, the
slow-roll condition is satisfied for $q$.  In this case, $q$ keeps its
initial value until the expansion rate of the universe becomes
comparable to the effective mass of the quintessence field.  This is
the case for the cosine-type quintessence models in the early
universe.

As the universe expands after the inflation, the horizon expands more
rapidly than the physical scale, and hence all the momentum scales
relevant to our discussion eventually enter the horizon.  Once the
scale enters the horizon, the spatial derivative term cannot be
neglected.  The energy density of such modes behaves like that of
relativistic matter, and hence it decreases as $a^{-4}$.  Thus, once
$k_{\rm phys}\sim H$ is realized, $\tilde{q}(k)$ decreases as
$a^{-1}$.

So far, we have seen that the perturbation in the quintessence
amplitude $q$ can be generated by several sources.  If $q$ is
non-vanishing, the energy density of the quintessence field
fluctuates.  The energy density, pressure, and momentum perturbations
of the quintessence are given by
\begin{eqnarray}
    \delta \rho_Q = \dot{\bar{Q}} \dot{q} + V' q,~~~
    \delta p_Q = \dot{\bar{Q}} \dot{q} - V' q,~~~
    (\rho_Q + p_Q) (v_Q)_i = 
    - \left( \frac{a_0}{a} \right) \dot{\bar{Q}}
    \frac{\partial q}{\partial x^i},
\end{eqnarray}
respectively.  These perturbations will affect the density
perturbation of the universe and may change the behavior of the CMB
anisotropy, as will be discussed in the following section.

\section{Numerical Results}
\label{sec:numerical}
\setcounter{equation}{0}

\subsection{Outline of the Analysis}

To study the CMB anisotropy in models with quintessence, we solve the
perturbed Einstein equation coupled to the equation of motion of the
quintessence field (as well as evolution equations of other components
like CDM, photon, baryon, and neutrinos).  Then, we compare the
theoretical prediction of the CMB anisotropy with observations.

As a first step, the total density and pressure perturbations are
derived by adding the quintessence contributions.  Notice that these
quantities affect the evolution of the metric perturbations.  Then, we
solve the evolution equations for the perturbations including the
quintessence contributions.  Evolution of $\bar{Q}(t)$ is
simultaneously solved since it affects the expansion rate and the
equation of state at each epoch.  In particular, for the perturbed
part of the equations, we modify the {\tt CMBFAST} package
\cite{cmbfast} to include the quintessence contributions.  Then, we
calculate the CMB anisotropy for the $l$-th multipole $C_l$, which is
defined as
\begin{eqnarray}
\left\langle 
\Delta T(\vec{x},\vec{\gamma})
\Delta T(\vec{x},\vec{\gamma}')
\right\rangle =
\frac{1}{4\pi} \sum_l (2l+1) C_l P_l 
(\vec{\gamma} \cdot \vec{\gamma}'),
\end{eqnarray}
where $\Delta T(\vec{x},\vec{\gamma})$ is the temperature fluctuation
of the CMB pointing to the direction $\vec{\gamma}$, and the average
is over the position $\vec{x}$.

Since we have not specified any particular model of inflation which
determines the normalization of the CMB anisotropy, over-all
normalization of $C_l$ is undetermined at this stage.  We treat the
normalization of the CMB anisotropy as a free parameter to make our
study model-independent.  Thus, in our analysis, the normalization of
the CMB anisotropy is fixed so that the theoretical prediction has the
best fit to the observational data.  We denote the actual CMB
anisotropy $C_l$ as
\begin{eqnarray}
C_l = N \bar{C}_l,
\end{eqnarray}
where $\bar{C}_l$ is the normalization-free CMB anisotropy and $N$ is
the normalization factor which minimizes the $\chi^2$ defined below.

Since accurate measurements of the CMB anisotropy have been performed
by COBE \cite{cobe}, BOOMERanG \cite{boomerang} and MAXIMA
\cite{maxima}, we can constrain the quintessence models by comparing
the theoretical prediction on $C_l$ with observations.  For this
purpose, we calculate the goodness-of-fit parameter $\chi^2=-2\ln L$,
where $L$ is the likelihood function.  (We call this parameter as
$\chi^2$, since it reduces to the usual $\chi^2$-variable for a
Gaussian.)  In our statistical analysis, following Ref.\ \cite{Bond},
we use the offset lognormal approximation to derive $\chi^2$ as
\begin{eqnarray}
\chi^2 = \sum_{BB'} (Z_B^{\rm th}-Z_B^{\rm obs})
M_{BB'}^Z (Z_{B'}^{\rm th}-Z_{B'}^{\rm obs}),
\label{chi^2}
\end{eqnarray}
where the summation is over the band powers obtained by COBE,
BOOMERanG and MAXIMA.  The quantity $Z_B^{\rm obs}$ contains
informations from the observations as
\begin{eqnarray}
Z_B^{\rm obs} &=& \ln (D_B + x_B),
\end{eqnarray}
where $D_B$ is the observed band power in $B$-th band and $x_B$ is the
offset correction, and $M_{BB'}^Z$ is given by
\begin{eqnarray}
M_{BB'}^Z = M_{BB'}^D (D_B + x_B) (D_{B'} + x_{B'}),
\label{M^Z}
\end{eqnarray}
where $M_{BB'}^D$ is the weight matrix for the band powers $D_B^2$.
(In Eq.\ (\ref{M^Z}), no summation over $B$ nor $B'$ is implied.)  In
addition, $Z_B^{\rm th}$ is written as
\begin{eqnarray}
Z_B^{\rm th} = \ln \left( N \sum_l u_\alpha f_{Bl} \bar{C}_l 
+ x_B \right),
\end{eqnarray}
where $f_{Bl}$ is the filter function and $u_\alpha$ is the
calibration parameter.\footnote{
In our analysis, we assume that the calibration uncertainty $u_\alpha$
has a flat distribution.  Therefore, $\chi^2$ given in Eq.\ 
(\ref{chi^2}) does not contain contributions from the calibration
uncertainties.}
In our analysis, we
neglect the calibration uncertainty in COBE and assume flat (top-hat)
distributions of $u_{\rm BOOMERanG}=1\pm 0.2$ and $u_{\rm MAXIMA}=1\pm
0.08$ \cite{jaffe}.  For a given set of $\bar{C}_l$, we vary $N$,
$u_{\rm BOOMERanG}$, and $u_{\rm MAXIMA}$ to minimize the $\chi^2$
parameter.  In our numerical calculations, we use {\tt RADPACK}
package \cite{radpack} to calculate $\chi^2$, which is based on 24,
12, and 10 band powers from COBE, BOOMERanG, and MAXIMA, respectively.

\subsection{Adiabatic Fluctuation}

Now, we are at the position to study the CMB anisotropy in models with
quintessence.  In this subsection, we consider the effects of the
adiabatic perturbation so we take $r_q=0$.

\begin{figure}[t]
    \centerline{\epsfxsize=0.75\textwidth\epsfbox{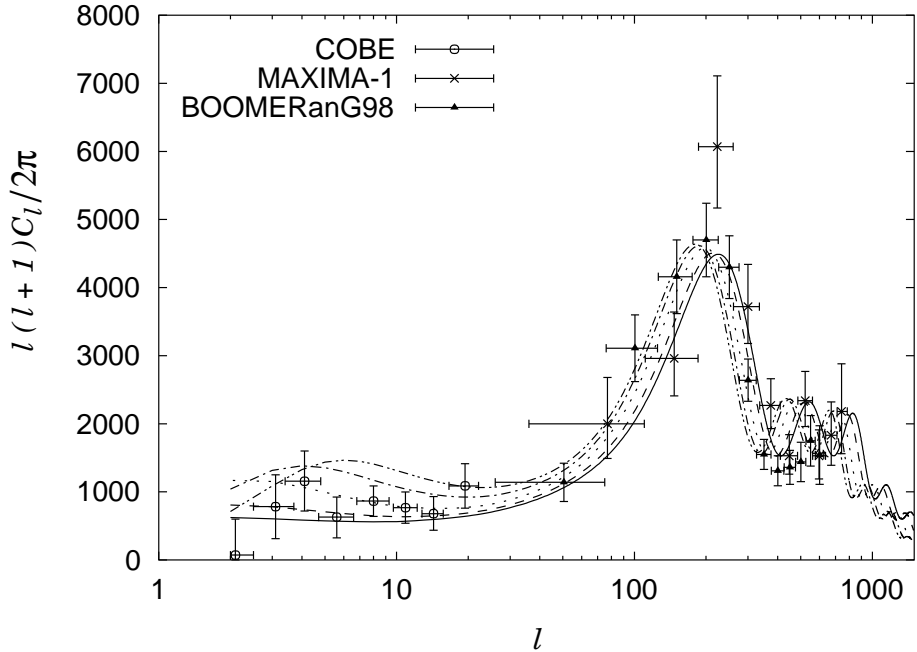}}
    \caption{The CMB angular power spectrum 
    $l(l+1)C_l/2\pi$ in models with cosine-type quintessence. The
    parameters we take here, $f_{Q}= 1.8 \times 10^{18}$ GeV, $\Lambda
    = 2.4 \times 10^{-12}$ GeV (dashed line), $\Lambda = 3.0 \times
    10^{-12}$ GeV (dotted line), $\Lambda = 4.0 \times 10^{-12}$ GeV
    (dash-dot line), $\Lambda = 5.0 \times 10^{-12}$ GeV (dash-dot-dot
    line). For comparison, we also show the cosmological constant case
    (solid line). Here, the cosmological parameters are taken to be
    $h=0.65$, $\Omega_{\rm m}=0.3$, $\Omega_{\rm b}h^{2}=0.019$, and
    the initial spectral index is $n=1$.  We also show the data points
    from COBE, BOOMERanG, and MAXIMA.  (For COBE, we use the reduced
    data set given in Ref.\ \cite{aph9702019}.)}
    \label{fig:Cl(adi)}
\end{figure}

In Fig.\ \ref{fig:Cl(adi)}, we first show the CMB angular power
spectrum.  Here, the cosmological parameters are taken to be $h=0.65$,
$\Omega_{\rm m}=0.3$, $\Omega_{\rm b}h^{2}=0.019$, and the initial
spectral index is $n=1$. We assume that there are no tensor mode
contributions.  The normalization factor $N$ is chosen such that
$\chi^2$ is minimized.

Some of the interesting features of the CMB angular power spectrum are
discussed in order.  First, let us consider the locations of the
acoustic peaks.  The locations of the peaks depend on two quantities,
the sound horizon at last scattering and the angular diameter distance
to the last scattering surface.  Approximately, the $l$-th multipole
picks up scales around $l \sim k r_{\theta}(\tau_{*}) $ where
$r_{\theta}(\tau_{*}) $ is the angular diameter distance to the last
scattering surface \cite{HuSug}.  The $n$-th peak in the temperature
power spectrum is located at the scale $k_n$ which satisfies
$k_nr_{s}(\tau_{*})=n\pi$, where $r_{s}(\tau_{*})$ is the sound
horizon at last scattering.  So the location of $n$-th peak in the $l$
space is estimated as
\begin{eqnarray}
    l_{n} \simeq \frac{r_\theta (\tau_*)}{r_{s}(\tau_{*})} n \pi.
\end{eqnarray}
Since the behavior of the cosine-type quintessence is almost the same
as that of the cosmological constant until very recently, the sound
horizon at last scattering is the same in both cases.  The angular
diameter distance in the quintessence model is, however, different
from that in the $\Lambda$CDM models.  Since the quintessence models
provides larger total energy density of the universe than the
$\Lambda$CDM models in the earlier epoch, the angular diameter
distance in the quintessence model becomes smaller than that in the
$\Lambda$CDM models.  As we can see from Fig.\ \ref{fig:Cl(adi)}, the
location of the peaks is shifted to lower multipole $l$ for the
quintessence models.  If we take a parameter in the oscillatory
region, this feature becomes more prominent.  As $\Lambda$ becomes
larger, more energy density exists in the early universe since some
fraction of the energy density of the quintessence damps away during
the oscillation.

Next let us consider the height of the acoustic peaks.  Since the
energy density of the quintessence becomes dominant when $z\sim O(1)$,
the late time integrated Sachs-Wolfe (ISW) effect enhances low
multipoles.  Such an enhancement may be more effective in the
quintessence models than in the $\Lambda$CDM models since, in the
quintessence case, the ``dark energy'' (i.e., the quintessence) may
dominate the universe earlier than in the $\Lambda$CDM case.  As a
result, with the COBE normalization, the height of the first peak
becomes lower.  On the contrary, since the quintessence becomes the
dominant component of the universe only at later epoch, pattern of the
acoustic oscillation before the recombination does not change compared
to $\Lambda$CDM models.  Therefore, ratios of the height of the first
peak to those of higher peaks are the same as $\Lambda$CDM models.

There is another point which should be addressed.  If we take a
parameter in the deep oscillatory region, we can see a bump in the low
multipole region.  At low multipoles, the late time ISW effect is
important for the temperature fluctuations in $\Lambda$CDM models or
in the quintessence models.  The ISW effect is originated from the
decay of the gravitational potential at the time of the quintessence
(or the cosmological constant) domination.  Before the quintessence
field starts to oscillate, the equation of state of the quintessence
field is almost $-1$, and hence this epoch is like the
cosmological-constant-dominated epoch.  After the quintessence starts
to oscillate, however, $\omega_Q$ approaches to 0 and hence the
quintessence field behaves like a matter component.  Then, the
gravitational potential $\Psi$ takes a constant value again, and the
ISW effect becomes ineffective.  Consequently, the bump in low
multipoles shows up.  Notice that the location of this bump
corresponds to the transition scale from the slow-roll epoch to the
oscillatory epoch.

\begin{figure}[t]
    \centerline
    {\epsfxsize=0.75\textwidth\epsfbox{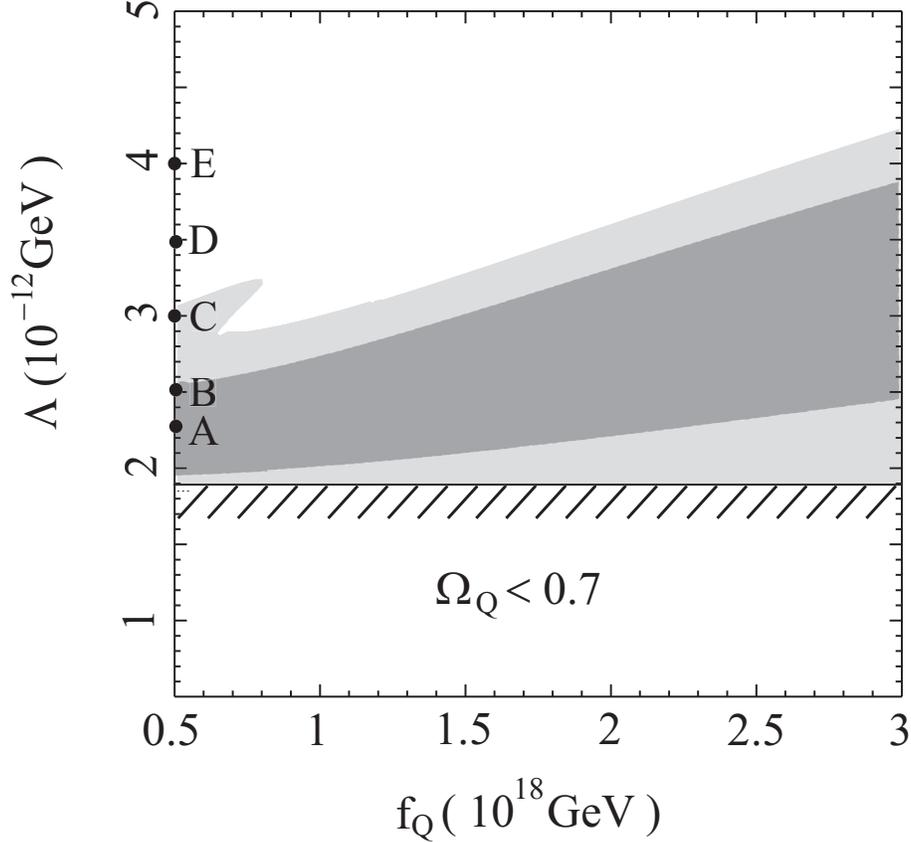}}
    \caption{Constraints on the parameters $\Lambda$ and $f_{Q}$.
    The lightly shaded regions are for $\chi^{2} \le 68.7$, and the
    darkly shaded region is for $\chi^{2} \le 60.5$. The cosmological
    parameters are taken to be $h=0.65$, $\Omega_{\rm m}=0.3$,
    $\Omega_{\rm b}h^{2}=0.019$, and the initial spectral index is
    $n=1$.  The Points A $-$ E will be used as representative points
    in the later discussion.}
    \label{fig:chi2}
\end{figure}

Now we discuss constraints on the cosine-type quintessence models from
the observations of COBE, BOOMERanG and MAXIMA.  Following the
prescription given in the previous subsection, we calculate the
goodness-of-fit parameter $\chi^2$ as a function of $f_Q$ and
$\Lambda$.  Based on this $\chi^2$ variable, constraints on the
parameter $f_Q$ and $\Lambda$ are shown in Fig.\ \ref{fig:chi2}.
Here, we show the region consistent with $\chi^{2}\leq 60.5 $ and
$\chi^{2}\leq 68.7$, which correspond to 95 \% and 99 \% C.L.\ allowed
region for the $\chi^{2}$-statistics with 44 degrees of freedom,
respectively.

When we take a parameter in the oscillatory regions, the quintessence
field becomes dominant component of the universe at earlier epoch.
Namely the late time ISW effect becomes large and the angular diameter
distance to the last scattering surface becomes smaller as mentioned
before.  Therefore, if $\Lambda$ is significantly large, the late time
ISW effect enhances angular power spectrum at low multipoles.  As a
result, the heights of the acoustic peaks are suppressed relative to
$C_l$ with small $l$.

\subsection{Isocurvature Mode}

Now, let us consider the effect of the isocurvature fluctuation which
may be generated in the very early universe (like during the
inflation).  

Before discussing the CMB anisotropy generated by the isocurvature
mode, let us first study the behavior of the fluctuation in the
quintessence amplitude $\tilde{q}$.  Since the isocurvature
contribution to $C_l$ does not interfere with the adiabatic one,
studying the purely isocurvature case is enough to understand the
effect of the isocurvature perturbation.

\begin{figure}[t]
    \centerline{\epsfxsize=0.75\textwidth\epsfbox{
       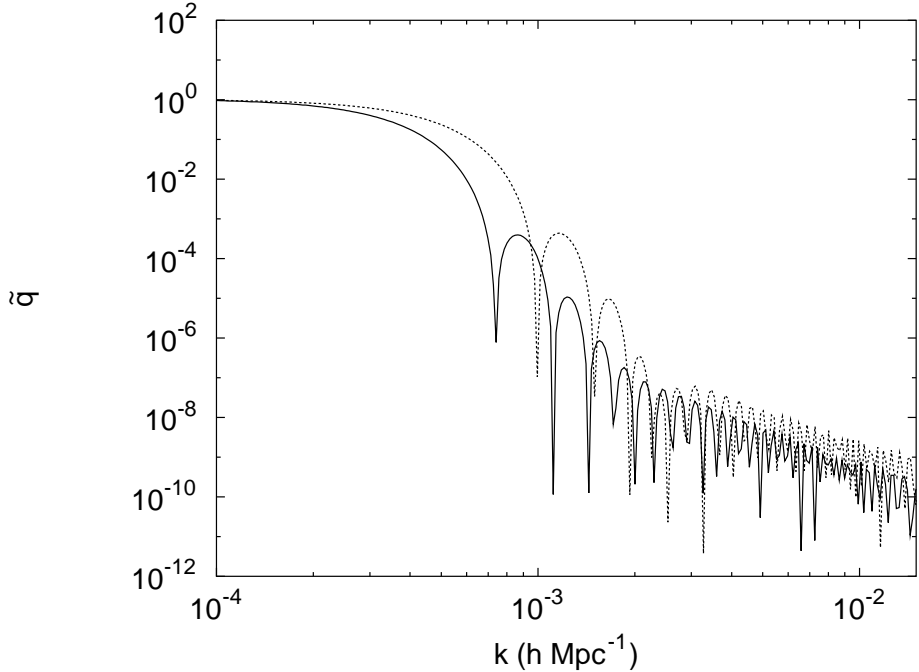}}
    \caption{$\tilde{q}$ as a function of $k$.  
      $\tilde{q}$ is normalized to unity at large scale (small $k$).
      Here, the model parameters are taken to be 
      $f_{Q}= 5.0 \times 10^{17}$ GeV and
      $\Lambda = 2.3 \times 10^{-12}$ GeV (solid line),
      $f_{Q}= 5.0 \times 10^{17}$ GeV and
      $\Lambda = 3.5 \times 10^{-12}$ GeV (dashed line) .}
    \label{fig:q(k)}
  \end{figure}

In Fig.\ \ref{fig:q(k)}, we show the scale dependence of $\tilde{q}$.
As one can see, $\tilde{q}(k)$ changes its behavior at the scale which
is slightly smaller than the present horizon scale: $\tilde{q}(k)$ is
almost constant for small $k$ while $\tilde{q}(k)$ is suppressed for
large $k$.  This behavior is understood as follows.  When the
slow-roll condition is satisfied for the quintessence, fluctuation for
the scale larger than the horizon scale takes its initial value.  On
the contrary, for $k_{\rm phys}\gtrsim H$, $\tilde{q}$ behaves like an
oscillator with frequency $\sim k_{\rm phys}$, and the fluctuation
red-shifts as the universe expands.  Due to the red shift, the
magnitude of $\tilde{q}$ is approximately proportional to $k^{-2}$ and
$k^{-1}$ for modes entering the horizon in the matter-dominated and
radiation-dominated universe, respectively.  If the slow-roll
condition is satisfied until the present universe, $\tilde{q}(k)$ for
$k_{\rm phys}\gtrsim H_0$ is suppressed.

If $m_{\rm eff}\sim\Lambda^2/f_Q\gtrsim H_0$, however, the
quintessence field may start to oscillate at some stage of the
universe.  (We call the horizon scale at this epoch as $1/k_{\rm
phys}^{\rm (osc)}$.)  Once the quintessence starts to oscillate with
frequency $m_{\rm eff}$, then all the modes universally damps as far
as $k_{\rm phys}\lesssim m_{\rm eff}$.  Thus, for $k_{\rm
phys}\lesssim k_{\rm phys}^{\rm (osc)}$, $\tilde{q}$ is independent of
$k$.  On the contrary, modes with $k_{\rm phys}\gtrsim k_{\rm
phys}^{\rm (osc)}$ acquire extra suppression as we discussed, and
$\tilde{q}$ for such a small scale becomes smaller.  In our analysis
we only consider cases where $1/k_{\rm phys}^{\rm (osc)}$ is close to
the present horizon scale.  In those cases, the isocurvature
fluctuation in the quintessence field dominantly affects $C_l$ with
small $l$, as we will see below.

\begin{figure}[t]
    \centerline{\epsfxsize=0.75\textwidth\epsfbox{
    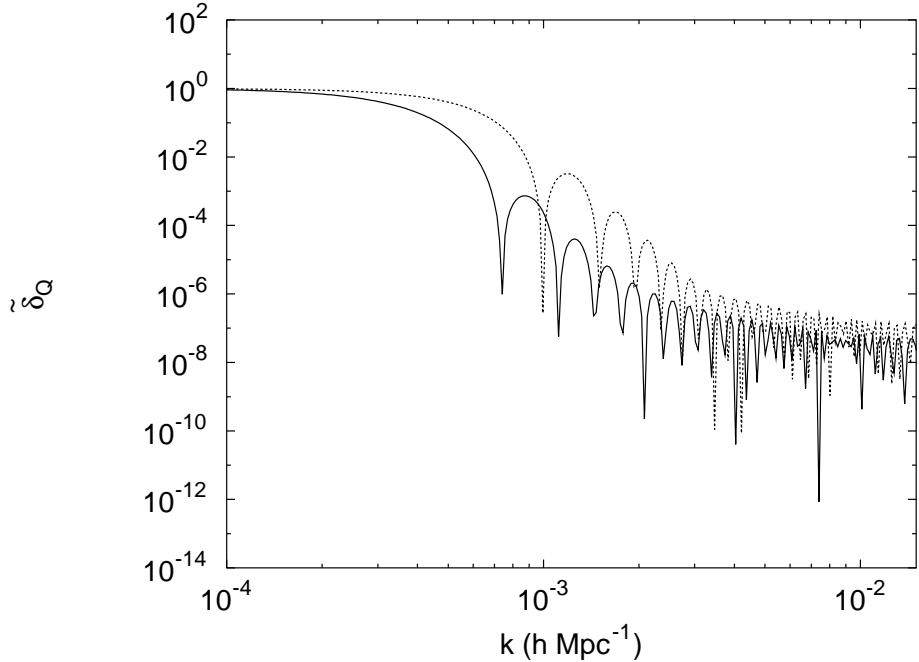}}
    \caption{$\tilde{\delta}_{Q}$ as a function of $k$. 
    $\tilde{\delta}_{Q}$ is normalized to unity at large scale (small
    $k$).  Cosmological and model parameters are the same as Fig.\ 
    \ref{fig:q(k)}.}
    \label{fig:D_Q}
\end{figure}

In Fig.\ \ref{fig:D_Q}, we also plot the scale dependence of the
energy density perturbation in the quintessence field normalized by
$\rho_Q$
\begin{eqnarray}
    \tilde{\delta}_Q =
    \frac{\delta\rho_Q}{\rho_Q}
    =\frac{\dot{\bar{Q}}\dot{\tilde{q}}+V'\tilde{q}}{\rho_Q}.
\end{eqnarray}
For the isocurvature contribution, the energy density perturbation is
dominated by that of the quintessence and hence it induces the metric
perturbation $\Psi$.  This becomes a source of the CMB anisotropy
through the Sachs-Wolfe effect.  When the slow-roll condition is
satisfied for the quintessence, $\tilde{\delta}_Q(k)$ is approximated
as
\begin{eqnarray}
    \tilde{\delta}_Q \simeq
    \frac{V'}{\rho_Q} 
    \left( -\frac{2\dot{\tilde{q}}}{9H} + \tilde{q} \right),
\end{eqnarray}
and it changes its behavior at the scale $k_{\rm phys}\sim H_0$ or at
$k_{\rm phys}\sim\Lambda^2/f_Q$, as a consequence of the scale
dependence of $\tilde{q}$.

\begin{table}
    \begin{center}
        \begin{tabular}{lccccc}
            \hline\hline
            {} & {$r_q=0$} & {$r_q=0.5$} & {$r_q=1$} 
            & {$r_q=1.5$} & {$r_q=2$} \\
            \hline
            {A} & {1.31} & {1.31} & {1.32} & {1.33} & {1.34} \\
            {B} & {1.45} & {1.46} & {1.48} & {1.51} & {1.56}  \\
            {C} & {1.15} & {1.26} & {1.61} & {2.18} & {2.98} \\
            {D} & {0.84} & {1.33} & {2.80} & {5.25} & {8.69} \\
            {E} & {0.62} & {1.93} & {5.86} & {12.41} & {21.58} \\
            \hline\hline
        \end{tabular}
        \caption{$C_2/C_{10}$ for several values of 
        $r_q$.  We take $f_Q=5\times 10^{17}\ {\rm GeV}$, and (A)
        $\Lambda=2.3\times 10^{-12}$ GeV, (B) $\Lambda=2.5\times
        10^{-12}$ GeV, (C) $\Lambda=3.0\times 10^{-12}$ GeV, (D)
        $\Lambda=3.5\times 10^{-12}$ GeV, and (E) $\Lambda=4.0\times
        10^{-12}$ GeV.  Notice that the Points A $-$ E are indicated
        in Fig.\ \ref{fig:chi2}.}
        \label{table:c2/c10}
    \end{center}
\end{table}

Now, we consider the CMB anisotropy in the case with the isocurvature
mode.  We calculate the CMB anisotropy for various cases, and in Table
\ref{table:c2/c10}, we show the quadrupole $C_2$ normalized by
$C_{10}$.

\begin{figure}
    \centerline{\epsfxsize=0.75\textwidth\epsfbox{
    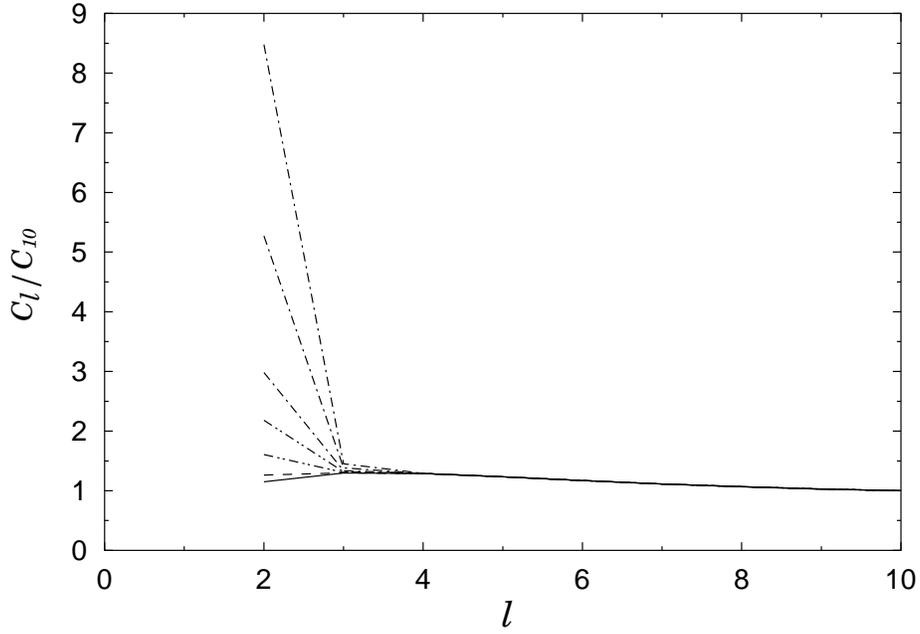}}
    \caption{$C_l/C_{10}$ for $r_q=0$, 0.5, 1, 1.5, 2, 3, and 4 from
    below with $f_Q=5\times 10^{17}\ {\rm GeV}$ and $\Lambda=3.0\times
    10^{-12}$ GeV.}
    \label{fig:C_2PointC}
\end{figure}

If we limit ourselves to the parameter region which is consistent with
the COBE, BOOMERanG, and MAXIMA observations with simple
scale-invariant primordial fluctuation, effect of the isocurvature
mode is quite small as far as $r_q\lesssim 1$.  This is because, in
such cases, there is a severe upper bound on $\Lambda$ to suppress the
late time ISW effect which enhances $C_l$ with small $l$.  As a
result, the quintessence field cannot dominate the universe when $z\gg
1$.  Then, the isocurvature fluctuation in the quintessence density
also becomes a minor effect until very recently. As one can see in
Table \ref{table:c2/c10}, for the best-fit value of $\Lambda$ (i.e.,
for the Point A given in Fig.\ \ref{fig:chi2}), the enhancement of
$C_2$ is about 2 \% even for $r_Q=2$.  If we consider larger value of
$\Lambda$, effect on $C_2$ is more enhanced.  For
$(f_Q,\Lambda)=(5\times 10^{17}\ {\rm GeV},3.0\times 10^{-12}\ {\rm
GeV})$ (i.e., for the Point C given in Fig.\ \ref{fig:chi2}, which is
allowed at 99 \% C.L.), we calculate the CMB angular power spectrum
and the result is given in Fig.\ \ref{fig:C_2PointC}.  In this case,
$C_2$ can be enhanced by the factor 2.6 if $r_Q=2$.

\begin{figure}
    \centerline{\epsfxsize=0.75\textwidth\epsfbox{
    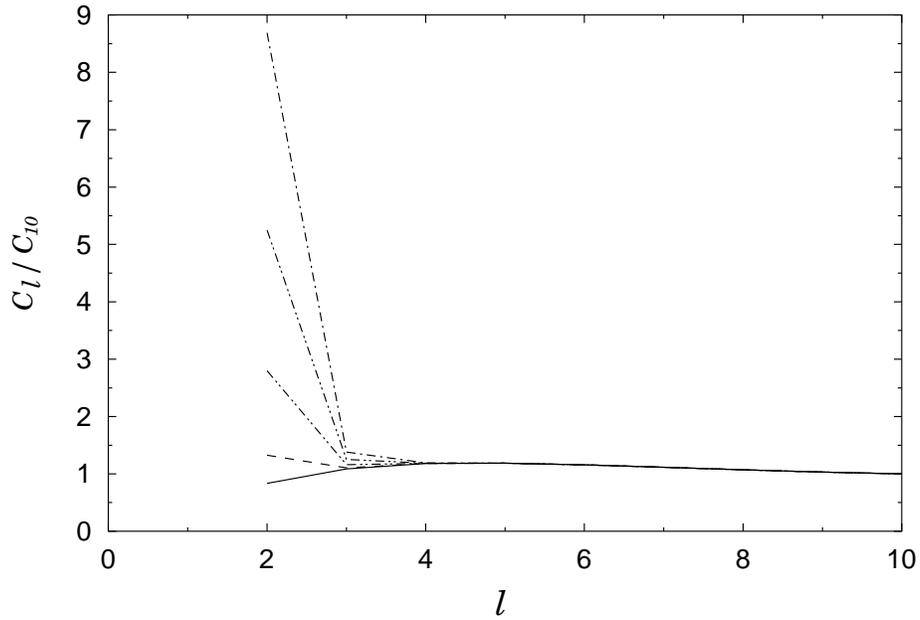}}
    \caption{$C_l/C_{10}$ for $r_q=0$, 0.5, 1, 1.5, and 2 from
    below with $f_Q=5\times 10^{17}\ {\rm GeV}$ and $\Lambda=3.5\times
    10^{-12}$ GeV.}
    \label{fig:C_2PointD}
\end{figure}

One should note that angular power spectrum of the CMB anisotropy
strongly depends on the primordial spectrum of the fluctuations.
Thus, the constraint on the $f_Q$ vs.\ $\Lambda$ plane is sensitive to
the scale-dependence of the primordial adiabatic fluctuation which is
determined by the model of the inflation.  Therefore, if we adopt a
possibility of a non-trivial scale dependence of the primordial
fluctuation, the constraint on the $f_Q$ vs.\ $\Lambda$ plane given in
the previous section may be relaxed or modified. If this is the case,
larger value of $\Lambda$ may be allowed and the energy density of the
quintessence field may become significant at earlier stage of the
universe.  Thus, we also consider such cases.  In Fig.\ 
\ref{fig:C_2PointD}, we plot $C_l$ normalized by $C_{10}$ for
$(f_Q,\Lambda)=(5\times 10^{17}\ {\rm GeV}, 3.5\times 10^{-12}\ {\rm
GeV})$. In addition, in Table \ref{table:c2/c10}, the ratio
$C_2/C_{10}$ is shown for the Points D and E.  As one can see, effect
on $C_2$ is much more significant than the previous case.

At low multipoles the uncertainty of the CMB data is dominated by
``cosmic variance.'' For $C_2$ which is mostly enhanced by the
isocurvature fluctuations, the cosmic variance gives $\sqrt{2/5}\simeq
60\ \%$ error to observation.\footnote{
The measurement of the CMB polarization toward many clusters would
allow some further reduction of the cosmic 
variance~\cite{Kamionkowski}.}
Thus, it is difficult to see the effect
of the isocurvature mode when $r_q\lesssim 1$ if the quintessence
field does not oscillate so much until the present epoch.  However, if
a large value of $\Lambda$ is possible, the isocurvature effects may
be detectable even with $r_q \gtrsim 0.5$, which is realized, for
example, in the chaotic inflation model with $V_{\rm
inf}\propto\chi^p$ with $p\gtrsim 8$ (see, for example, Point E in
Table \ref{table:c2/c10}).

\section{Conclusions and Discussion}
\label{sec:conclusion}
\setcounter{equation}{0}

We have studied the CMB anisotropies produced by cosine-type
quintessence models.  In particular, effects of the adiabatic and
isocurvature fluctuations have been both discussed.

For purely adiabatic fluctuations with scale invariant spectrum, the
existence of the quintessence suppresses the relative height of the
first acoustic peak of the angular power spectrum compared with the
$\Lambda$CDM case.  This is because, in the quintessence models, the
``dark energy'' due to the quintessence may dominate the universe
earlier than the cosmological constant case, and hence the late time
ISW effect becomes more effective.  As a result, the CMB anisotropy
for large angular scale is more enhanced, which relatively suppresses
the height of the acoustic peaks.  Because of this effect, we have
seen that the CMB data from COBE, BOOMERanG and MAXIMA have imposed
the stringent constraint on the model parameters of the quintessence
models.  We have also seen that the location of the the first acoustic
peak shifts to lower multipole $l$ compared with $\Lambda$CDM models.

In the case of the cosine-type quintessence models, the quintessence
field has a negligible effective mass during inflation and hence its
amplitude may acquire sizable fluctuation due to the quantum
fluctuation in the de Sitter background.  Such a fluctuation becomes
isocurvature fluctuation.  We have shown that the isocurvature
fluctuations have significant effects on the CMB angular power
spectrum at low multipoles in some parameter space, which may be
detectable in future satellite experiments.  This signal may be used
to test the cosine-type tracker model, combining with the global shape
of the CMB angular power spectrum.  In particular, enhancements at low
multipoles do not exist in the minimal $\Lambda$CDM models.

In the tracker-type models, the CMB angular power spectrum may be also
affected by the isocurvature mode \cite{aph0103244}.  Contrary to the
cosine-type case, however, tracker field starts to evolve in the early
universe and hence its effective mass is as large as the expansion
rate of the universe.  As a result, fluctuation in the tracker field
damps while the tracker field follows the attractor solution.  In
addition, in the tracker case, the flatness of the tracker potential
is not guaranteed by any symmetry during the inflation.  (Notice that
for the cosine-type case some symmetry may keep the flatness of the
potential.)  As a result, in the de Sitter background, the tracker
field may acquire effective mass as large as $H_{\rm inf}$ during the
inflation.  This is the case for, for example, quintessence models
based on supergravity with minimal K\"ahler potential.  If such a
large mass exists, fluctuation generated during the inflation is
damped, as can be seen in Eq.\ (\ref{q(k)}).  These effects may
drastically suppress the signal of the isocurvature mode.  Detailed
analysis of the tracker models will be given in elsewhere
\cite{KawMorTak}.

Since very accurate measurements of the CMB anisotropy are expected in
the near future, in particular by the MAP \cite{MAP} and PLANCK
\cite{PLANCK} experiments, we will have more stringent constraints on
the quintessence models as well as on the $\Lambda$CDM models.
Importantly, the quintessence and $\Lambda$CDM models may have
different predictions on the shape of the CMB angular power spectrum,
and some of the models may be confirmed or excluded once better
observations of the CMB anisotropy become available.  In particular,
in the cosine-type quintessence models, we have seen that the
isocurvature perturbation may enhance the CMB angular power spectrum
at low multipoles (in particular, $C_2$), which may be an interesting
signal of the cosine-type quintessence models.

{\sl Acknowledgment:} The authors would like to thank J. Hwang for
useful comment.  M.K. thanks N. Sugiyama for useful discussion.  This
work is supported by the Grant-in-Aid for Scientific Research from the
Ministry of Education, Science, Sports, and Culture of Japan, No.\ 
12047201, No.\ 13740138 and Priority Area ``Supersymmetry and Unified
Theory of Elementary Particles'' (No.\ 707).

\end{document}